\documentclass[letterpaper, 10 pt, journal]{IEEEtran}  % Comment this line out if you need a4paper

\IEEEoverridecommandlockouts                              % This command is only needed if 
\usepackage{epsfig} % for postscript graphics files
\usepackage{mathptmx} % assumes new font selection scheme installed

\usepackage{times}

\usepackage{times}
\usepackage{bm}
\usepackage{url}
\usepackage{tikz}
\usepackage{cite}
\usepackage{dsfont}
\usepackage{color} 
\usepackage{upgreek}
\usepackage{footnote}
\usepackage{verbatim}
\usepackage{mathrsfs} 
\usepackage{graphicx}
\usepackage{paralist}
\usepackage{algorithm}
\usepackage{algorithmicx}
\usepackage{algpseudocode}
\algrenewcommand\textproc{} % to change function names to lower case
\usepackage{amssymb, amsmath, times}
\usepackage[pagebackref=false,breaklinks=true,colorlinks=true, citecolor=magenta,linkcolor=cyan,urlcolor=blue,bookmarks=true]{hyperref}

\DeclareGraphicsExtensions{.pdf,.jpeg,.png,.eps,.ai}
{}

\title{\LARGE \bf
	 A Second-Order  Reachable Sets Computation Scheme via a  Cauchy-Type Variational Hamilton-Jacobi-Isaacs Equation.}

\author{Lekan Molu$^\star$, Ian Abraham$^\ddagger$, and Sylvia Herbert$^\dagger.$    
	\thanks{\footnotesize $^\star$Microsoft Research, 300 Lafayette Street, New York, NY10012, USA. $^\ddagger$Department of Mechanical Engineering, Yale University, New Haven, Conn, USA, $^\dagger$Department of Mechanical and Aerospace Engineering, UC San Diego, La Jolla, CA, USA.  
		{\tt\small \{lekanmolu@microsoft.com, ian.abraham@yale.edu, sherbert@eng.ucsd.edu
			\}}. }
}	
\begin{document}
		
\maketitle
\thispagestyle{empty}
\pagestyle{empty}
		
\definecolor{light-blue}{rgb}{0.30,0.35,1}
\definecolor{light-green}{rgb}{0.20,0.49,.85}
\definecolor{purple}{rgb}{0.70,0.69,.2}

\newcommand{\lb}[1]{\textcolor{light-blue}{#1}}
\newcommand{\rev}[1]{\textcolor{red}{#1}}

\renewcommand{\figureautorefname}{Fig.}
\renewcommand{\sectionautorefname}{$\S$}
\renewcommand{\equationautorefname}{equation}
\renewcommand{\subsectionautorefname}{$\S$}
\renewcommand{\chapterautorefname}{Chapter}

% FYA
\newcommand{\cmt}[1]{{\footnotesize\textcolor{red}{#1}}}%{#2}
\newcommand{\todo}[1]{\textcolor{cyan}{TO-DO: #1}}
\newcommand{\review}[1]{\noindent\textcolor{red}{$\rightarrow$ #1}}
%Text commands
\newcounter{mnote}
\newcommand{\marginote}[1]{\addtocounter{mnote}{1}\marginpar{\themnote. \scriptsize #1}}
\setcounter{mnote}{0}
\newcommand{\ie}{i.e.\ }
\newcommand{\eg}{e.g.\ }
\newcommand{\cf}{cf.\ }
\newcommand{\yes}{\checkmark}
\newcommand{\no}{\ding{55}}

%Reference commands
\newcommand{\flabel}[1]{\label{fig:#1}}
\newcommand{\seclabel}[1]{\label{sec:#1}}
\newcommand{\tlabel}[1]{\label{tab:#1}}
\newcommand{\elabel}[1]{\label{eq:#1}}
\newcommand{\alabel}[1]{\label{alg:#1}}
\newcommand{\fref}[1]{\cref{fig:#1}}
\newcommand{\sref}[1]{\cref{sec:#1}}
\newcommand{\tref}[1]{\cref{tab:#1}}
\newcommand{\eref}[1]{\cref{eq:#1}}
\newcommand{\aref}[1]{\cref{alg:#1}}

\newcommand{\bull}[1]{$\bullet$ #1}
\newcommand{\argmax}{\text{argmax}}
\newcommand{\argmin}{\text{argmin}}
\newcommand{\mc}[1]{\mathcal{#1}}
\newcommand{\bb}[1]{\mathbb{#1}}

\newcommand{\shmargin}[2]{{\color{magenta}#1}\marginpar{\color{magenta}\raggedright\footnotesize #2}}
\newcommand{\shnote}[1]%
{\textcolor{magenta}{SH: #1}}
\newcommand{\lmnote}[1]%
{\textcolor{orange}{LM: #1}}

\def\tidx{t}
%\def\comment
%\def\value{V}
% from https://www.cs.jhu.edu/~jason/advice/write-the-paper-first.html
\newcommand{\Note}[1]{}
\renewcommand{\Note}[1]{\hl{[#1]}}  % comment out this

\def\kau{\mc{K}}
\def\particle{\bm{x}}
\def\materialresponse{\bm{G}}
\def\orthoggroup{{\textit{SO}}(3)}
\def\liegroup{{\textit{SE}}(3)}
\def\liealgebra{\mathfrak{se}(3)}
\def\identity{\bm{I}}
\newcommand{\trace}[1]{\textbf{tr}(#1)}

\def\rot{{R}}
\def\rthree{\bb{R}^3}
\def\reline{\bb{R}}
\def\targetset{\mathcal{L}}
\def\traj{\xi}
\def\ren{\bb{R}^n}
\def\skew{S}
\def\jacob{J}
\def\state{\bm{x}}
\def\statex{x}
\def\statey{y}
\def\statez{z}
\def\hot{h.o.t.\ }
\def\lhs{l.h.s.\ }
\def\rhs{r.h.s.\ }
\def\identity{I}
\def\costdiff{\mathbf{\tilde{V}}}
\def\pursuer{\bm{P}}
\def\evader{\bm{E}}
\def\gain{\bm{k}}
\def\control{\bm{u}}
\def\disturb{\bm{v}}
\def\switchcurve{\bm{\gamma}}
\def\valuefunc{\bm{V}}
\def\valueterm{\bm{g}}
\def\lpspace{L^2({\mc{S}}; \mc{F})}
\def\lpdual{L^2({\mc{S}}; \breve{\mc{F}})}
\def\valuetensor{\mathds{V}}
\def\wtensor{\mathds{W}}
\def\valuecore{\mathds{V}^c}
\def\hamfunc{\bm{H}}
\def\hamtensor{\mathds{H}}
\def\uppervalue{\bm{V}^+}
\def\lowervalue{\bm{V}^-}
\def\upperham{\bm{H}^+}
\def\lowerham{\bm{H}^-}
\def\hilbertparam{\bm{\phi}}
\def\hilbertcoeff{\bm{\psi}}
\def\hilbertparamspace{\bm{\Phi}}
\def\hilbertcoeffspace{\bm{\Psi}}
\def\hilbertspace{\mathcal{F}}
\def\hilbertdual{\mathcal{S}}
\def\reducedbasis{\Xi_r}
\def\basis{\mathbf{e}}
\def\openset{\Omega}
\def\spatialdomain{\Omega}
\def\timeinterval{I}
\def\liederi{L}
\begin{abstract}
	Motivated by the scalability limitations of a popular Eulerian variational Hamilton-Jacobi-Isaacs (HJI) computational method for backward reachability analysis -- essentially providing a least restrictive controller in   problems that involve state or input constraints under a worst-possible disturbance, 
	we introduce a second-order approximation algorithm for computing the zero sublevel sets of the variational HJI equation. By continuously approximating the value functions of all possible trajectories within a dynamical system's state space, under sufficient HJI partial differential equation regularity and continuity conditions throughout the state space, we show that with proper line search for step size adjustments and state feedback control under the worst-possible disturbance, we can compute the state set that are reachable within a prescribed verification time bound.
\end{abstract}
\section{Introduction}

At issue is returning \textit{point sets} for dynamical systems' behavioral evolution (or trajectory) throughout a state space. Such sets must satisfy \textit{input or state constraints} in a \textit{least restrictive sense} under a \textit{worst-possible disturbance}~\cite{LygerosReachability}, and \textit{over a time period}. This is a classic reachability theory problem. In this letter, we shall provide an approximation scheme for constructing the \textit{discriminating kernel}  for a controller's ``constraints-satisfaction" design specifications problem.  Our study is motivated by the exponential computational requirement of Eulerian methods~\cite{Mitchell2005, LevelSetsBook}. Ours is an extension of differential dynamic programming methods in Bolza-like objective functions inspired by~\cite{JacobsonMayne, Bryson}, and ~\cite{JacobsonThesis}'s second-order variational methods for solving general optimal control problems. In emphasis, we focus on pursuit-evasion  games~\cite{Isaacs1965}, where in an iterative dynamic game fashion~\cite{iDG}, each agent's strategy depends on its opponent's control law as agents locally approximate successive trajectories and their values that emanate from a state space. We adopt a formal treatment, indicating how to resolve these problems computationally.%, which utilize grid discretization and are well-known to scale exponentially for practical problems~\cite{LevelSetsBook}.   

%A basic characteristic of a discrete, continuous, or an hybrid control system is to determine the point set within its state space that are \textit{reachable} by the choice of a control input over a time interval.

\textit{Reachable sets} can be analyzed in a 
\begin{inparaenum}[(i)]
	\item \textit{forward} sense, where system trajectories are examined to determine if they enter certain states from an \textit{initial set};
	\item \textit{backward} sense, where system trajectories are examined to determine if they enter certain \textit{target sets};
	\item \textit{reach set} sense, in which they are examined to see if states reach a set at a \textit{particular time}; or
	\item \textit{reach tube} sense, in which they are examined to determine that they reach a set \textit{during a time interval}. %\cite{Mitchell2007Comparing}.	
\end{inparaenum} 

We focus on the backward construction. This consists in avoiding an unsafe state set under the worst-possible disturbance within a given time-bound. The state sets of a backward reachable problem constitute the \textit{discriminating kernel} that is ``safety-preserving" for a finite-time horizon control problem. This discriminating kernel's boundary satisfies a generalized Isaac's equation \ie if the state $\state$ is within the set's boundary, but not the target's boundary, then for a co-state $p$ that is an exterior proximal normal at $\state$, $\hamfunc(\state, p) \le 0$, where $\hamfunc(\state, p)$ is the problem's Hamiltonian. The set associated with the kernel computed in a backward reachability  framework is termed the \textit{backward reachable set} or \textit{BRS}. 
% For an infinite-time horizon control problem, they constitute the \textit{largest robust controlled-invariant set}. In the absence of a disturbance, the resulting kernel is said to satisfy \textit{viability constraints}. 

%BRS  are popularly analyzed as a game of two vehicles with non-stochastic dynamics~\cite{Mitchell2005, Mitchell2020}. 
Eulerian methods~\cite{SethianLSBook, LevelSetsBook} resolve the BRS as the zero-level set of an implicitly-defined value function on a state space. Constructed as an initial value problem for a \textit{Cauchy-type} Hamilton-Jacobi-Isaacs (HJI) partial differential equation (P.D.E.)~\cite{Mitchell2005, MitchellLSToolbox2007}, the BRS in one dimension is equivalent to the conservation of momentum equation~\cite{LevelSetsBook}. For a multidimensional problem, by resolving the P.D.E. on a dimension-by-dimension basis in a consistent and monotone fashion~\cite{Crandall1984Approx}, a numerically precise and accurate solution to the HJ P.D.E can be  determined~\cite{SethianLSBook, LevelSetsBook}.  However, as state dimensions increase, spacetime discretization methods become impractical owing to their exponential complexity. Contrary to Eulerian methods, our approximation method does not require state space discretization and hence reduces the exponential complexity to polynomial time.

\textbf{Contributions}: We describe a computational scheme, provide a summary of the complexity and convergence behavior for the overapproximated BRS using~\cite{Mitchell2005}'s standard variational Hamilton-Jacobi equation.  Iteratively approximating nonlinear trajectories about ``near" local paths and under positive-definiteness requirement of the stagewise cost function's second-derivatives (in control terms), we compute the extrema of the nominal state and control-disturbance policy pair's cost to find a \textit{cost improvement} per iteration. Trajectories are updated until we  sweep all possible initial conditions into the space of \textit{all trajectory costs}. Within the bounds here set, the zero isocontour of all ``unionized costs" of all trajectories swept along paths that assure cost improvement per iteration of the algorithm constitute the zero level set. This work is the first to systematically provide a polynomial time complexity computational scheme for backward reachable sets to our knowledge.

The body of this letter is structured as follows: \autoref{sec:notations} describes the notations used . In \autoref{sec:back}, we introduce methods that will enable us  formulate our proposal in \autoref{sec:methods}. We describe the computational and complexity analysis in \autoref{sec:procedure}.  % rational incremental decomposition scheme that provides approximation guarantees on regions of the state space where approximate HJ control laws are valid as well as provide a rational analysis for high-dimensional verification of nonlinear systems with guarantees.  %\todo{Add more stuff that encompasses the problems we address in this paper here.}     
\section{Notations, Terminology, and Assumptions}
\label{sec:notations}

We employ standard vector-matrix notations throughout. Conventions: small  Latin and Greek letters are scalars; in bold-font they are vectors. Exceptions: Players in a differential game e.g. $\pursuer, \, \evader$, are individual entities. All vectors are column-stacked. Capital Greek and Calligraphic letters are sets. 
The scalar product of vectors $\state \text{ and } \bm{y}$ is written as $\langle \state, \bm{y}\rangle$. Arbitrary real variables e.g., $t, t_0, t_f, \tau, T$ denote \textit{time}. % while fixed, ordered values of $t$ are denoted as $t_0 \le t \le t_f$. The transpose of a matrix, $\bm{X}$, is written as $\bm{X}^T$. 
 For a scalar function $\valuefunc$, $\valuefunc_{\state}$ is the gradient vector, and $\valuefunc_{\state \state}$ is the jacobian matrix. 

\noindent \textbf{Differential Games}: 
At issue are conflicting objectives between two players in pursuit-evasion games with each player being a pursuer ($\bm{P}$) or an evader ($\bm{E}$).
The set of all \textit{strategies} executed by $\pursuer$ (resp. $\evader$) during a game (beginning at a time $t$) is denoted as $\mc{B}(t)$ (resp. $\mc{A}(t)$).

\noindent \textbf{System Description}: For the dynamical system
\begin{align}
	\dot{\state}(t) &= f(t, \state(t), \bm{u}(t), \bm{v}(t)), \,\, \state(0) = \state_0,  \,\, -T \le t \le 0,
	\label{eq:sys_dyn}
\end{align}
\noindent where the state $\state(t)$ evolves from some initial negative time $-T$ to a final time $0$. 
\noindent \textbf{Assumptions}: 
The flow field $f(t, \cdot, \cdot, \cdot)$ and $\state(t)$ are  bounded and Lipschitz continuous for fixed controls $\control(t)$ and $\disturb(t)$; $f(t, \cdot, \cdot, \cdot)$ possesses $\mathcal{C}^2$ continuity in $\state$\footnote{This bounded Lipschitz continuity property assures uniqueness of the system response $\state(t)$ to controls $\bm{u}(t)$ and $\bm{v}(t)$~\cite{Evans1984}.}. We take $\state(t)$ to belong in the open set $\openset \subset \bb{R}^n$ and the pair $(\bm{x}, t)$ as the system's \textit{phase}. The Cartesian product of $\openset$ and the space $T =\reline^1$ of all time values is termed the \textit{phase space}, $\openset \times T$. The closure of $\openset$ is denoted $\bar{\openset}$ and we let $\delta \openset$ denote the boundary of $\openset$. Unless otherwise stated, vectors $\bm{u}(t) \in \bb{R}^{n_u}$ and $\bm{v}(t) \in \bb{R}^{n_v}$ are reserved for admissible control (resp. disturbance) at time $t$ for some  $n_u, \, n_v > 0$. Controls $\bm{u}(t)$ (resp. $\bm{v}(t)$) are piecewise continuous in $t$, if for each $t$, $\bm{u} \in \mathcal{U}$ (resp. $\bm{v} \in \mathcal{V}$) and $\mathcal{U}(\text{resp. } \mc{V})$ are Lebesgue measurable and compact sets. At all times, any of $\control(t)$ or $\disturb(t)$ will be under the influence of a \textit{player} such that the motion of $\state(t)$ will be influenced by the will of that player. When a control law or value function is optimal, it shall be signified by an asterisk superscript e.g. $\control^\star$. 
For the phase space $\left(\Omega \times T\right)$,  the set of all controls for players $\pursuer$ and $\evader$ are respectively drawn from
\begin{align}
	\mathcal{\bar{U}} &\equiv \{\bm{u}: [-T, 0] \rightarrow \mathcal{U} | \bm{u} \text{ measurable}, \, \mathcal{U} \subset \bb{R}^{n_u} \}, \\
	\mathcal{\bar{V}} &\equiv \{\bm{v}: [-T, 0] \rightarrow \mathcal{V} | \bm{v} \text{ measurable},  \,\mathcal{V} \subset \bb{R}^{n_v}\},
\end{align}
with $\mathcal{U}, \, \mathcal{V}$ being compact.

\textbf{Existence and Uniqueness of Value and Trajectories}: 
For any admissible control-disturbance pair $(\bm{u}(t), \bm{v}(t))$ and initial phase $(\state, -T)$, given a pursuer's \textit{non-anticipative strategy}, the game admits a value~\cite{Elliott1974} and there exists a unique trajectory  $\bm{\xi}(t)$ such that the motion of \eqref{eq:sys_dyn} passing through phase $(\bm{x}, -T)$ under the action of control $\bm{u}(t)$, and disturbance $\bm{v}(t)$, and observed at a time $t$ afterwards, given by,
\begin{align}
	\bm{\xi}(t) = \bm{\xi}(t; -T, \state, \bm{u}(\cdot), \bm{v}(\cdot))
	\label{eq:HJ_traj}
\end{align}
satisfies \eqref{eq:sys_dyn} almost everywhere (a.e.)~\cite{Mitchell2005}. Note that under feedback strategies, the game may not admit a value necessarily~\cite{Cardaliaguet97Nonsmooth}; however, if the target sets are initialized such that there is no overlap of states between players, and that the evader(s) are not captured by the pursuer(s) at the start of the game, then one can guarantee a pursuit winning strategy~\cite[Assumption 2.1]{MatchingCapture}.
%\section{Background} 
%\label{sec:prelim} 
%% 
%\todo{potentially add in the stuff about HJI here ?}

%\shnote{Time is a bit scrambled up in this section.  need to set straight}
\section{Background.}
\label{sec:back}

Reachable sets in the context of dynamic programming and two person games is here introduced. We restrict attention to  computing the backward reachable sets of a dynamical system. We establish the viscosity solution P.D.E. to the terminal HJI P.D.E, and then describe the formulation of the BRS and backward reachable tube (BRT). Let us enquire.

\subsection{The Backward Reachable (Target) Set}
\noindent For any optimal control problem, a value function is constructed based on a user-defined  optimal cost that is bounded and uniformly continuous for any input phase $(\state, -T)$ e.g. reach goal at the end of a time horizon i.e., %.  In reachability analysis, typically this is defined using a terminal cost function $g(\cdot): \bb{R}^n \rightarrow \bb{R}$ that satisfies 
	\begin{align}
		| g(0; \state) | &\le k, \,\,
		| g(0; \state) - g(t; \hat{\state}) \mid \le k | \state - \hat{\state} \mid
	\end{align}
for constant $k$ and all 
$-T \le t \le 0$, $\hat{\state}, \, \state \in \reline^n$.  The set
\begin{align}
	\mathcal{L}_0 = \{ \state \in \bar{\Omega} \,|\, g(0; \state) \le 0 \},
	\label{eq:target_set}
\end{align}
is the \textit{target set} in the phase space $\openset \times \mathbb{R}$ (proof in \cite{Mitchell2005}). This target set can represent the failure set (to avoid) or a goal set (to reach) in the state space. %Note that the target set, $\mathcal{L}_0$, is a closed subset of $\ren$ and is in the closure of $\openset$. Typically $\mathcal{L}_0$ is user-defined, and $g(x)$ is a signed distance function, that is negative inside the target set and positive elsewhere.

\subsection{The Backward Reachable Tube}
Backward reachability analysis seeks to capture all conditions under which trajectories of the system may enter a user-defined target set cf. \eqref{eq:target_set}.  This could be desirable in goal-regions of the state (safe sets) or undesirable state configurations (unsafe sets). For a target set construction problem, a differential game's (lower) value is equivalent to a solution of \eqref{eq:sys_dyn} for $\bm{u}(t)$ and $\bm{v}(t) = \beta[\control](t)$ i.e.,
\begin{align}
	&\valuefunc(\state, t) = \inf_{\beta \in \mathcal{B}(t)} \sup_{\bm{u} \in \mathcal{U}(t)} 
	\min_{t\in[-T,0]}
	g\left(0; t, \bm{x}, \control(\cdot), \disturb(\cdot)\right). %
	\label{eq:value_lower}
\end{align}
Optimal trajectories emanating from an initial phase $(\state,-T)$ where the value function is non-negative will maintain non-negative cost over an entire time horizon, thereby avoiding the target set and vice versa. % Optimal trajectories from initial phases where the value function is negative will enter the target set at some point within the time horizon.  
For the safety problem setup in \eqref{eq:value_lower}, we can define the corresponding \textit{robustly controlled backward reachable tube} %for $\tau \in [-T, 0]$
 as the closure of the open set
\begin{align}
	\mathcal{L}([\tau, 0], \mathcal{L}_0) &= \{\state \in \openset \,| \, \exists \, \beta \in \mathcal{\bar{V}}(t) \,  \forall \, \bm{u} \in \mathcal{U}(t), \exists \, \tau \in [-T, 0], \nonumber \\
	& \qquad  \qquad \bm{\xi}(\bar{t})%\left(t; \state_0, t_0, \bm{u}(\cdot), \beta[\bm{u}](\cdot) \right)
	\in  \mathcal{L}_0 \}. %\,\bar{t} \in \left[-T, 0\right].
	\label{eq:rcbrt}
\end{align}
Read: The set of states from which the strategies of $\pursuer$ and for all controls of $\evader$ imply that we \textit{reach and remain in the target set} in the interval $[-T, 0]$.  Following Lemma 2 of \cite{Mitchell2005}, the states in the reachable set admit the following properties w.r.t the value function $\valuefunc$:
%
%\begin{subequations}
	\begin{align}
		\state(t)\in \mathcal{L}(\cdot) \implies \valuefunc(\state, t) \le 0, %\\
		\valuefunc(\state, t) \le 0 \implies \state(t) \in \mathcal{L}(\cdot). 
		\label{eq:reachables} 
	\end{align}
%\end{subequations}
%
Player  $\pursuer$ is minimizing (the game's termination time c.f. \eqref{eq:target_set}), seeking to    drive system trajectories into the unsafe set; and  $\evader$ is maximizing (the game's termination time) \ie is seeking to avoid the unsafe  set-\footnote{For the goal-satisfaction (or \textit{liveness}) problem setups, the strategies are reversed and the backward reachable tube are the states from which the evader $\evader$ can successfully reach the target set despite worst-case efforts of the pursuer $\pursuer$.}.

\subsection{The Terminal HJI Value Function for Reachability}
%Rather than computing the minimum cost for every possible trajectory of the system, in safety analysis it is sufficient to consider the minimum cost under optimal behavior from both players.  The optimal behavior of each player depends on whether the target set represents a goal or a failure set. %For a safety (avoiding a failure set) problem setup, the evader $\evader$ is seeking to maximize the minimum cost (keeping the system out of the target set) and the pursuer $\pursuer$ seeks to minimize it. 

The \textit{non-anticipative strategy} used in level set methods follows from ~\cite{Evans1984} viz., suppose that the pursuer's strategy starting at a time $-T$, is $\beta: \mathcal{\bar{U}} \rightarrow \mathcal{\bar{V}}$. Suppose further that $\beta$ is provided for each $-T \le \tau \le 0$ and $\bm{u}, \hat{\bm{u}} \in \mathcal{\bar{U}}$. Then,
\begin{align}
	\begin{cases}
		&\bm{u}(\bar{t}) = \hat{\bm{u}}(\bar{t}) \,\, \text{ a.e. on } -T \le \bar{t}  \le \tau,  \\ 
		&\text{implies } \beta[\bm{u}](\bar{t}) = \beta[\hat{\bm{u}}](\bar{t}) \,\, \text{ a.e. on } -T \le \bar{t}  \le \tau.
	\end{cases}
\end{align} 

In our work, however, we use feedback strategies in an iterative dynamic game framework while initializing the target sets to ensure the existence of a value as discussed in \S \ref{sec:notations}. %Computing the value function is in general challenging and non-convex. %Additionally, the value function is hardly smooth throughout the state space, so it lacks classical solutions even for smooth Hamiltonian and boundary conditions. 
It is well-known that a differential game's value function admits a ``viscosity" (generalized)  solution~\cite{Lions1982, Crandall1983viscosity} of the associated HJ-Isaacs (HJI) PDE %which are \textit{locally Lipschitz} in $\openset \times [-T, 0]$, and with at most first-order partial derivatives in the Hamiltonian.  Equation \eqref{eq:value_lower} admits the viscosity solution
given by
	\begin{align}
		\frac{\partial \valuefunc}{\partial t}{(\bm{x}, t)} & + \min \{0, \hamfunc (t; \bm{x}, p) \} = 0, \,\, %\\
		V(\bm{x},0) = g(\bm{x})
		\label{eq:lower_hji_pdes}
	\end{align}
\noindent where the vector field $p\,(\equiv\valuefunc_{\state})$ is known in terms of the game's terminal conditions so that the overall game is akin to a two-point boundary-value problem; and the Hamiltonian  $\hamfunc(t; \bm{x}, \bm{u}, \bm{v},\valuefunc_{\state})$ is defined as 
\begin{align}
	\hamfunc(t; \bm{x}, \bm{u}, \bm{v},\valuefunc_{\state})= \max_{u \in \mathcal{U}} \min_{v \in \mathcal{V}} \, \langle f(t; \state, \bm{u}, \bm{v}), \valuefunc_{\state}  \rangle.
\end{align}
Equation \eqref{eq:value_lower} only allows the game to determine that a trajectory belongs in the target set at exactly time zero. That is, the evader can chase a trajectory into the target set and escape it before reaching the final time, $0$.  The $\min$ operation between the scalar $0$ and the Hamiltonian allows the pursuer to ``freeze" trajectories that may be under the evader's willpower when the evader tries to evolve such trajectories outside of the target set. 
For more details on the construction of this P.D.E., see \cite{Mitchell2005}. %In Mitchell's construction, the letter, t
\section{Successive Approximation Scheme}
\label{sec:methods}
Throughout this section, all feasible trajectories $\{\state_i, \state_{i+1}, \cdots, \state_{n}\} \in \openset$  are scheduled for iterative second-order expansion along the nonlinear state variations $\delta \state_i$ and about nominal states $\bar{\state}_i$; we derive the variational linear differential equation form of \eqref{eq:lower_hji_pdes} and describe the integration scheme of the backward and forward steps of standard DDP on the variational HJI equation. In what follows, for ease of notation we drop the subscripts that denote the respective trajectories. For clarity's sake, we shall drop the subscripts and derive the optimal value for a generic trajectory $\state$.

\subsection{Local Approximations  to Nonlinear Trajectories}

Suppose that system \eqref{eq:sys_dyn} is controllable everywhere on an interval $[-T, 0]$ along a local trajectory $\state_r$ generated by admissible control disturbance pair $\control_r$, $\disturb_r$ starting from an initial phase $(\state_r, -T)$ so that
\begin{align}
	\dot{\state}_r(\tau) &= f(t; \state_r(\tau), \control_r(\tau), \disturb_r(\tau)). %; %\quad \, \state_r(-T) = \state_{r_{(-T)}}.
\end{align}
First, we apply local controls $\control_r(t) \, \text{ and } \disturb_r(t)$ on \eqref{eq:sys_dyn} so that the nominal value  is $\valuefunc(t; \cdot, \cdot, \cdot)$ for a resulting nominal state $\state_r(\tau)$; $\tau \in \left[-T, 0\right]$. System dynamics that describe variations from the nonlinear system \cf \eqref{eq:sys_dyn} with state and control pairs $\delta \state(t), \, \delta \control(t), \,\delta \disturb(t)$ respectively\footnote{Note that $\delta \state(t), \, \delta \control(t), \,\text{ and } \delta \disturb(t)$ are respectively measured with respect to $\state(t), \control(t), \disturb(t)$ and are not necessarily small.} can then be written as
\begin{subequations}
	\begin{align}
		\state(t) &= \state_r(t) + \delta\state(t), \,\, 	\control(t) = \control_r(t) + \delta\control(t), \\
		\disturb(t) &= \disturb_r(t) + \delta\disturb(t), \,\, t \in \left[-T, 0\right].
	\end{align}
	\label{eq:variations}
\end{subequations}
Abusing notation, we drop the templated time arguments in the variations \eqref{eq:variations} so that the canonical problem is now 
%
%\begin{subequations}
	\begin{align}
		\dfrac{d}{dt}\left(\state_r + \delta \state\right) &= f(t; \state_r + \delta \state, \control_r + \delta \control, \disturb_r + \delta \disturb), \,  \\
	 \state_r(-T) 	&+ \delta \state(-T) = \state(-T),
	 \label{eq:canonical_variation}
	\end{align}
whose terminal value admits the optimal form (see \eqref{eq:lower_hji_pdes}):
\begin{align}
	\begin{split}
		-\frac{\partial \valuefunc^\ast}{\partial t}(\state_r + \delta \state, t)
		&= 
		\min \left\{\bm{0},  
		\max_{\delta \control \in \mathcal{U}} \, \min_{\delta \disturb \in \mathcal{V}} \left\langle f(t; \state_r + \delta \state, \right. \right. \\
		&  \left. \left.  \control_r + \delta \control, \disturb_r + \delta \disturb), \dfrac{\partial \valuefunc^\ast}{\partial \state}\left(\state_r + \delta \state, t \right) \right\rangle \right\},
	\end{split} \nonumber \\
	\valuefunc^\ast(\state_r + \delta \state, 0) &= g(0; \state_r(0) + \delta \state (0));
	\label{eq:canonical_value}
\end{align}
and state trajectory
\begin{align}
	\bm{\xi}(t) = \bm{\xi}(t; -T, \state_r + \delta\state, 	\control+\delta\control,\disturb+\delta\disturb), \quad t \in \left[-T, 0\right].
\end{align}
For $-T < t \le 0$ and a $\tau \in [t, 0]$, let the \textit{optimal cost}  for using the optimal control $\control^\star(\tau) = \control_r(\tau) + \delta \control^\star(\tau)$ and optimal disturbance $\disturb^\star(\tau) = \disturb_r(\tau) + \delta \disturb^\star(\tau)$ be $\valuefunc^\star(\state_r, \tau)$; and the \textit{nominal cost} for using $\control_r(\tau)$ be $\valuefunc_r(\state_r, t)$, so that on the phase $\left(\state_r, t\right)$, we have
\begin{align}
	\costdiff^\star(\state_r, t) =\valuefunc^\star(\state_r, t) -\valuefunc_r(\state_r, t).
	\label{eq:cost_diff}
\end{align}
Next, we expand $\valuefunc^\ast$ in \eqref{eq:canonical_value} under sufficient regularity assumptions.
\subsection{Power Series Expansion Scheme}
\label{subsec:main_result}
	 Suppose that the optimal terminal cost, $\valuefunc^\ast(\cdot)$, is sufficiently smooth to allow a power series expansion in the state variation $\delta \state$ about the nominal state, $\state_r$, then we must have
	\begin{align}
		\valuefunc^\star(\state_r + \delta \state, t) &=
		\costdiff^\star(\state_r, t) + \valuefunc_r(\state_r, t) +  \left\langle\valuefunc_{\state}^\ast(\state_r, t), \delta \state \right\rangle  
		\nonumber \\
		& \quad+ \dfrac{1}{2} \left\langle \delta \state, \valuefunc^\star_{\state\state}(\state_r, t) \delta\state \right\rangle   + \text{h.o.t in } \delta \state,
		\label{eq:volterra_expand}
	\end{align}
	where h.o.t. signifies higher order terms. The smoothness assumption is necessary for admitting the linear differential equation expansion of \eqref{eq:volterra_expand}.  This expansion is consistent with \textbf{differential dynamic programming} schemes~\cite{JacobsonThesis, JacobsonMayne, DenhamDDP}. 
	
	Observe: If $\delta \state$ is not constrained to be small, \eqref{eq:volterra_expand} may require huge memory for storage owing to the large dimensionality of terms beyond second order. However,
	\begin{inparaenum}[(i)]
		\item if $\state_r$ is constrained to be sufficiently close to $\state$, the state variation  $\delta \state$ will be small, resulting in $\state \approx \state_r$  \cf \eqref{eq:variations}\footnote{\label{ftnt:smallx}A scheme to keep the variation $\delta \state$ small is discussed in section \ref{sec:procedure}. Ultimately, the $\delta \control, \delta \disturb$ terms will be quadratic in $\delta \state$ if we neglect h.o.t.};
		\item for small $\delta \state$,  the expansion of \eqref{eq:volterra_expand}  becomes consistent with second-order methods where quadratic terms in $\delta \state$ dominate higher-order terms.
	\end{inparaenum}
	Hence, we can avoid the infinite data storage requirement by truncating the expansion in \eqref{eq:volterra_expand} at second-order terms in $\delta \state$. This will incur an $O(\delta \state^3)$ approximation error, affording us realizable control laws that can be executed on the system \eqref{eq:sys_dyn}.  Thus, we rewrite \eqref{eq:volterra_expand} as
	\begin{align}
		\valuefunc^\star(\state_r + \delta \state, t) &=
		\costdiff^\star(\state_r, t) + \valuefunc_r(\state_r, t) +  \left\langle\valuefunc_{\state}^\ast(\state_r, t), \delta \state \right\rangle  
		\nonumber \\
		& \quad+ \dfrac{1}{2} \left\langle \delta \state, \valuefunc^\star_{\state\state}(\state_r, t) \delta\state \right\rangle.
		\label{eq:val_rom_expand}
	\end{align}
	Denote by $\valuefunc_{\state}^\star\left(\state_r + \delta \state, t \right)$ the optimal value of the co-state $\frac{\partial\valuefunc^\star}{\partial \state}\left(\state_r + \delta \state, t \right)$ on the r.h.s of \eqref{eq:canonical_value}. Then, expanding up to second order we find that
	\begin{align}
		\valuefunc_{\state}^\star\left(\state_r + \delta \state, t \right) =  \valuefunc^\star_{\state} \left(\state_r , t \right) + \langle \valuefunc^\star_{\state\state}\left(\state_r , t \right), \delta \state \rangle,
		\label{eq:co_state_expand}
	\end{align}
	where we have again omitted the quadratic term $\valuefunc^\star_{\state\state\state} \delta \state \delta \state$ owing to the foregoing reason. 
	Substituting \eqref{eq:val_rom_expand} and \eqref{eq:co_state_expand} into \eqref{eq:canonical_value}, abusing notation by dropping the templated phase arguments, we find that %(recall that $\valuefunc_r(\state_r,t)$ is the cost for using the nominal control $\control_r(\tau), \, \tau \in [t, 0]$ for $-T < t \le 0$)
	\begin{align}
		\begin{split} 
			-\frac{\partial \costdiff^\ast}{\partial t} -\frac{\partial \valuefunc_r}{\partial t} - \left\langle \dfrac{\partial\valuefunc_{\state}^\ast}{\partial t}, \delta\state \right \rangle -  \dfrac{1}{2} \left\langle \delta \state, \dfrac{\partial \valuefunc_{\state\state}^\ast }{\partial t}\delta\state \right \rangle &=  \\
			\min \left\{0,  
			\max_{\delta \control \in \mc{U}} \, \min_{\delta \disturb \in \mc{V}} \left\langle f^T(t; \state_r + \delta \state, \control_r + \delta \control,  \disturb_r + \delta \disturb), \right. \right. \\
			\left. \left. \valuefunc^\ast_{\state}  + \valuefunc^\ast_{\state\state}\, \delta \state \right\rangle \right\}. 
		\end{split}
		\label{eq:ROM_HJI_Proof}
	\end{align}
	From \eqref{eq:cost_diff}, we may write
	\begin{subequations}
		\begin{align}
			&\frac{d}{dt}\left(\valuefunc_r + \costdiff^\ast\right) = \dfrac{\partial}{\partial t}\left(\valuefunc_r + \costdiff^\ast\right) + \left\langle f^T(t; \state_r, \control_r, \disturb_r), \valuefunc_{\state}^\ast \right \rangle \label{eq:opt_val_diff}\\
			\dot{\valuefunc}_{\state}^\ast &= \dfrac{\partial \valuefunc_{\state\state}^\ast}{\partial t} + \langle f^T(t; \state_r, \control_r, \disturb_r),  \valuefunc_{\state\state}^\ast \rangle, \,\, \dot{\valuefunc}_{\state\state}^\ast = \dfrac{\partial \valuefunc_{\state\state}^\ast}{\partial t}.
		\end{align}
		\label{eq:reduced_canonical}
	\end{subequations}
Therefore, \eqref{eq:ROM_HJI_Proof} in light of \eqref{eq:reduced_canonical} becomes 
\begin{align}
	\begin{split} 
		-\frac{\partial \costdiff^\ast}{\partial t} -\frac{\partial \valuefunc_r}{\partial t} - \left\langle \dfrac{\partial\valuefunc^\ast_{\state}}{\partial t}, \delta\state \right \rangle -  \dfrac{1}{2} \left\langle \delta \state, \dfrac{\partial \valuefunc^\ast_{\state\state} }{\partial t}\delta\state \right \rangle &=  \\
		\min \left\{\bm{0},  
		\max_{\delta \control \in \mc{U}} \, \min_{\delta \disturb \in \mc{V}} \left[\hamfunc(t; \state_r + \delta \state, \control_r + \delta \control, \disturb_r + \delta \disturb, \valuefunc_{\state}^\ast) + \right. \right. \\
		\left. \left.  \left\langle \valuefunc_{\state\state}^\ast\, \delta \state, f(t; \state_r + \delta \state, \control_r + \delta \control,  \disturb_r + \delta \disturb)  \right\rangle\right] \right\}
	\end{split}
\label{eq:rob_expand}
\end{align}
where $\hamfunc(t; \state, \control, \disturb, \valuefunc_{\state}^\ast) = \langle \valuefunc_{\state}^\ast, f(t; \state, \control, \disturb) \rangle$. 

\subsection{Variational HJI Linear Differential Equation}
\label{subsec:var_diff_eq}
Let the respective maximizing and minimizing control-disturbance pair when $\state=\state_r$ at time $t$ be
\begin{align}
	\control^\ast &= \control_r+\delta \control^\ast, \,\,
	\disturb^\ast = \disturb_r+\delta \disturb^\ast.
	\label{eq:bkwd_controls}
\end{align} 
Given the variation $\delta \state$ on the r.h.s. of  \eqref{eq:rob_expand}, write the respective maximizing evader's and minimizing pursuer's controls for state $\state=\state_r+\delta \state$ 
\begin{align}
	\control&=\control_r+\delta \control^\ast +\delta \control, \,\,
	\disturb=\disturb_r+\delta \disturb^\ast +\delta \disturb.
	\label{eq:fwd_controls}
\end{align} 
Note here that $\delta \control$ and $\delta \disturb$ are as yet to be determined.  	
Henceforward, we adopt the abbreviations $\hamfunc=\hamfunc(t; \state_r+\delta \state, \control_r+\delta \control, \disturb_r+\delta \disturb, \valuefunc_{\state}^\ast)$ and $f=f(t; \state_r+\delta \state, \control_r+\delta \control, \disturb_r+\delta \disturb)$. Expanding the r.h.s. of \eqref{eq:rob_expand} about $\state_r, \control^\ast, \disturb^\ast$ at time $t$ up to second-order as before, we have
\begin{align}
	\begin{split} 
		\min \left\{\bm{0},  
		\max_{\delta \control \in \mc{U}} \, \min_{\delta \disturb \in \mc{V}} \left[\hamfunc + \left\langle \hamfunc_{\state}  + \valuefunc^\ast_{\state \state} f, \delta \state\right\rangle + \langle \hamfunc_{\control }, \delta \control \rangle 	+ 
		\right. \right. \\ \left. \left. 
	 \langle \hamfunc_{\disturb }, \delta \disturb \rangle + \langle \delta \control, (\hamfunc_{\control  \state} + f_{\control}^T \valuefunc_{\state \state}^\ast) \delta \state  \rangle + \left\langle \delta \disturb, (\hamfunc_{\disturb  \state} + f_{\disturb}^T \valuefunc_{\state \state}^\ast) \delta \state  \right\rangle  \right.\right. \\
		\left. \left. 
		+ \dfrac{1}{2}\left\langle\delta \control, \hamfunc_{\control  \disturb} \delta \disturb + \dfrac{1}{2}\left\langle\delta \disturb, \hamfunc_{\disturb  \control} \delta \control \right\rangle \right\rangle +  \dfrac{1}{2} \left\langle \delta \control, \hamfunc_{\control  \control} \delta \control \right\rangle   \right. \right. \\
		\left. \left.
		+  \dfrac{1}{2} \left\langle \delta \disturb, \hamfunc_{\disturb  \disturb} \delta \disturb \right\rangle +    \dfrac{1}{2} \left\langle \delta \state, \left(\hamfunc_{\state \state} + f_{\state}^T \valuefunc_{\state \state}^\ast  + \valuefunc_{\state \state}^\ast f_{\state} \right) \delta \state \right\rangle \right] \right\}. 
	\end{split}
\label{eq:rob_rhs}
`\end{align}
When capture occurs \ie when $\evader$'s separation from $\pursuer$ becomes less than a pre-specified (capture) radius, we must have 
\begin{align}
	\hamfunc_{\control}(t; \state_r, \control^\star, \disturb, \valuefunc_{\state}) = 0; \,\hamfunc_{\disturb}(t; \state_r, \control, \disturb^\star, \valuefunc_{\state}) = 0
	\label{eq:saddle_funcs}
\end{align}                                                                 %
%where $\control_r^\star$ and $\disturb_r^\star$ respectively represent the optimal control laws for both players at time
for  $t \in \left(-T, 0\right]$. Seeking variational  feedback controllers of the form: $\delta \control = \gain_{\control} \delta \state, \, \delta \disturb = \gain_{\disturb} \delta \state$ and putting \eqref{eq:saddle_funcs} into \eqref{eq:rob_rhs} we may write 
\begin{subequations}
	\begin{align}
		 %&\hamfunc_{\control} + 
		 &\hamfunc_{\control  \control} \delta \control  + \left(\hamfunc_{\control  \state} + f_{\control}^T \valuefunc_{\state \state}^\ast\right) \delta \state + \dfrac{1}{2}\left(\hamfunc_{\control  \disturb}+\hamfunc_{\disturb \control  }^T\right) \delta \disturb = 0,  \\
		%
		 %&\hamfunc_{\disturb} + 
		 &\hamfunc_{\disturb  \disturb} \delta \disturb  + \left(\hamfunc_{\disturb  \state} + f_{\disturb}^T \valuefunc_{\state \state}^\ast\right) \delta \state + \dfrac{1}{2}\left(\hamfunc_{\disturb\control} + \hamfunc_{\control \disturb}^T \right)\delta \control = 0
		\label{eq:ham_zero}
	\end{align}
\end{subequations}
%Employing \eqref{eq:saddle_funcs}, and comparing terms in  equation \eqref{eq:ham_zero} to the introduced feedback form, 
so that evaluating \eqref{eq:ham_zero} at $\state_r$, $\control^\ast$, and $\disturb^\ast$ we find that
\begin{align}
	\gain_{\control} &= -\hamfunc_{\control  \control}^{-1} \left[ \hamfunc_{\control  \disturb}\gain_{\disturb}  + \left(\hamfunc_{\control  \state} + f_{\control}^T \valuefunc_{\state \state}^\ast\right) \right], \text{ and that } \nonumber \\
	\gain_{\disturb} &= - \hamfunc_{\disturb  \disturb}^{-1} \left[\hamfunc_{\disturb  \control}  \gain_{\control} +  \left(\hamfunc_{\disturb  \state} + f_{\disturb}^T \valuefunc_{\state \state}^\ast\right)\right]. 
	\label{eq:gains}
\end{align}        
In equation \eqref{eq:gains}, the two players' control strategies are interdependent: the optimal control law for the pursuer is to choose a control strategy that is an optimal response to the evader's control input. This is akin to a Newton-Raphson scheme which ``correctly" estimates the maximizing $\delta \control^\ast$ and and minimizing $\delta \disturb^\ast$ provided that $|\control-\control^\ast|, |\disturb-\disturb^\ast|<\epsilon \text{ for } \epsilon>0$ with the added requirement that $\hamfunc_{\control  \control}$ and $\hamfunc_{\disturb  \disturb}$ are  positive-definite. Upon substitution,  \eqref{eq:rob_rhs} satisfies
\begin{align}
\begin{split} 
	\min \left\{\bm{0}, \hamfunc + \left\langle \hamfunc_{\state}  + \valuefunc_{\state \state}^\ast f, \delta \state \right\rangle +  \frac{1}{2} \left\langle \delta \state,
	\right.\right. \\ \left.\left.
   \left(\hamfunc_{\state \state} + f_{\state}^T \valuefunc_{\state \state}^\ast  + \valuefunc_{\state \state}^\ast f_{\state}  + \gain_{\control}^T \hamfunc_{\control  \control}  \gain_{\control} 
	+ \gain_{\disturb}^T \hamfunc_{\disturb  \disturb}  \gain_{\disturb}\right) \delta \state \right\rangle\right\}. 
\end{split}
\label{eq:rob_ham_analytical}
\end{align}
We discard terms involving the pairs $(\delta \control, \,\delta \state)$ and $(\delta \disturb, \,\delta \state)$ beyond second order. Comparing  \eqref{eq:rob_ham_analytical} to the \lhs of \eqref{eq:rob_expand}, we find that
\begin{subequations}
	\begin{align}
		%\begin{split}
			&- \dfrac{\partial \tilde{\valuefunc}^\ast}{\partial t}-\dfrac{\partial \valuefunc_r}{\partial t} = \min\{\bm{0}, \hamfunc\}, %
			\\
			-\dfrac{\partial \valuefunc_{\state}^\ast}{\partial t} &= \min \left\{\bm{0}, \hamfunc_{\state} + \valuefunc_{\state\state}^\ast \, f\right\}, \\
			%
			%-  \dfrac{\partial \valuefunc_{\state \state}^\ast}{\partial t}
			-\dot{\valuefunc_{\state \state}^\ast} 
			&= \min \left\{\bm{0}, \hamfunc_{\state \state} + f_{\state}^T\valuefunc_{\state\state}^\ast + \valuefunc_{\state\state}^\ast f_{\state} 
			%\right. \nonumber \\ \left.
			%\nonumber \\%
			%\quad \quad  
			+ \gain_{\control}^T \hamfunc_{\control  \control}  \gain_{\control} +  \gain_{\disturb}^T \hamfunc_{\disturb  \disturb}  \gain_{\disturb} \right\}. \nonumber
		%\end{split}
	\end{align}
	\label{eq:variational_approx}
\end{subequations}
Furthermore, comparing \eqref{eq:variational_approx} with \eqref{eq:reduced_canonical}, and using the first-order necessary condition for optimality \cf \eqref{eq:saddle_funcs}, we have
 implies 
\begin{subequations}
	\begin{align}
		- \dot{ \tilde{\valuefunc}}^\ast &= \min\{\bm{0}, \hamfunc
	- \hamfunc(t; \state_r, \control_r, \disturb_r, \valuefunc_{\state}^\ast) 	\} \\
		-\dot{\valuefunc}_{\state}^\ast &= \min \left\{0, \hamfunc_{\state} + \valuefunc_{\state\state}^\ast \left(f-f(t; \state_r, \control_r, \disturb_r)\right) \right\} \\
		-\dot{\valuefunc_{\state \state}^\ast}  &= \min \left\{\bm{0}, \hamfunc_{\state \state}+ f_{\state}^T\valuefunc_{\state\state}^\ast + \valuefunc_{\state\state}^\ast f_{\state}  
		+ \gain_{\control}^T \hamfunc_{\control  \control}  \gain_{\control} 
		%\right. \\ \left. 
		+  \gain_{\disturb}^T \hamfunc_{\disturb  \disturb}  \gain_{\disturb} \right\} \nonumber
	\end{align}
	\label{eq:saddle_reduce}
\end{subequations}
where every quantity in \eqref{eq:saddle_reduce} is evaluated at $t; \state_r, \control^\ast, \disturb^\ast$\footnote{The ``$^\ast$" sign implies that the optimal $\control^\ast=\control_r+\delta \control^\ast$
	and $\disturb^\ast=\disturb_r + \delta \disturb^\ast$ are used -- essentially the optimal control-disturbance pair for $\state = \state_r + \delta \state$.}.  Note that equation \eqref{eq:saddle_reduce} signifies the  terms to be computed in the typical \textit{backward pass} of DDP-like algorithms namely,  start at the final time $t=0$ and proceed backwards until $-T$; whilst storing the locally linear control gains \ie \eqref{eq:gains} at every step of the backward pass to keep the necessary conditions of optimality. In a \textit{forward pass}, the following new controls 
\begin{subequations}
	\begin{align}
		\control(\tau) &= \control^\ast(\tau) +  \gain_{\control} \delta \state(\tau), \\
		\disturb(\tau) &= \disturb^\ast(\tau) +  \gain_{\disturb} \delta \state(\tau), \, \tau \in \left[-t, 0\right], \, t>T
	\end{align}
\label{eq:feedback_control}
\end{subequations}
are updated proceeding forward in time from $-T$ to $0$ \footnote{Note that gains $\gain_{\control}, \,\gain_{\disturb}$ are only used in computing feedback controllers for the discriminating kernel in a backward reachability setting.}. 

The linear differential variational HJI equation for \eqref{eq:lower_hji_pdes} thus becomes (for sufficiently small $\delta \state$): 
\begin{align}
	\begin{split} 
			-\frac{\partial \hat{\costdiff}}{\partial t} -\frac{\partial \valuefunc_r}{\partial t} - \left\langle \dfrac{\partial\hat{\valuefunc}_{\state}}{\partial t}, \delta\state \right \rangle -  \dfrac{1}{2} \left\langle \delta \state, \dfrac{\partial \hat{\valuefunc}{\state\state} }{\partial t}\delta\state \right \rangle =
			\min \left\{\bm{0}, 
		\right. \\ \left.
			\left[\hamfunc(t; \state_r, \control^\ast, \disturb^\ast, \hat{\valuefunc}_{\state}) + \left\langle \hamfunc_{\state}  + \hat{\valuefunc}_{\state \state} f, \delta \state\right\rangle 
			+ \langle \hamfunc_{\control }, \gain_{\control}\delta \state \rangle
		\right. \right. \\ \left. \left. 
		+\langle \hamfunc_{\disturb }, \gain_{\disturb} \delta \state \rangle \right.\right. \\ \left. \left. 
		 + \langle \gain_{\control}\delta \state, (\hamfunc_{\control  \state} + f_{\control}^T \hat{\valuefunc}_{\state \state}) \delta \state  \rangle + \left\langle  \gain_{\disturb} \delta \state, (\hamfunc_{\disturb  \state} + f_{\disturb}^T \hat{\valuefunc}_{\state \state}) \delta \state  \right\rangle  \right. \right. \\		\left. \left.
	+ \dfrac{1}{2}\left\langle \gain_{\control}\delta \state, \hamfunc_{\control  \disturb} \delta \disturb + \dfrac{1}{2}\left\langle\delta \disturb, \hamfunc_{\disturb  \control} \gain_{\control}\delta \state \right\rangle \right\rangle 	+  \dfrac{1}{2} \left\langle \gain_{\control}\delta \state, \hamfunc_{\control  \control} \gain_{\control}\delta \state \right\rangle   
		\right. \right. \\		\left. \left.
		 +  \dfrac{1}{2} \left\langle  \gain_{\disturb} \delta \state, \hamfunc_{\disturb  \disturb}  \gain_{\disturb} \delta \state \right\rangle + \dfrac{1}{2} \left\langle \delta \state, \left(\hamfunc_{\state \state} + f_{\state}^T \hat{\valuefunc}_{\state \state}  + \hat{\valuefunc}_{\state \state} f_{\state} \right) \delta \state \right\rangle \right] 
		\right\}. 
	\end{split}
	\label{eq:variational_approx_fwd}
\end{align}
where the hat terms indicate that \eqref{eq:feedback_control} is used. \textit{Note that  we have discarded the $\valuefunc_{\state \state \state}$ terms in our derivations. We also do away with the application of the penalizing $\epsilon>0$-term on the variational controls in \eqref{eq:feedback_control} as proposed in \cite{JacobsonThesis} in lieu of standard step-size adjustment mechanisms to ensure a sufficiently small $\delta \state$.}

In a similar spirit to \eqref{eq:saddle_reduce}, we have 
\begin{subequations}
	\begin{align}
		- \dot{ \hat{\valuefunc}}^\ast &= \min\{\bm{0}, \hamfunc
		- \hamfunc(t; \state_r, \control_r, \disturb_r, \hat{\valuefunc}_{\state}^\ast) 	\} \\
		-\dot{\hat{\valuefunc}}^\ast_{\state} &= \min \left\{0, \hamfunc_{\state} + \hat{\valuefunc}^\ast_{\state\state} \left(f-f(t; \state_r, \control_r, \disturb_r)\right) \right\} \\
		%
		%-  \dfrac{\partial \hat{\valuefunc}^\ast_{\state \state}}{\partial t} 
		-\dot{ \hat{\valuefunc}}^\ast_{\state \state}&= \min \left\{\bm{0}, \hamfunc_{\state \state}+ f_{\state}^T \hat{\valuefunc}^\ast_{\state\state} + \hat{\valuefunc}^\ast_{\state\state} f_{\state}  
		+ \gain_{\control}^T \hamfunc_{\control  \control}  \gain_{\control} 
	%	\right. \\ \left. 
		+  \gain_{\disturb}^T \hamfunc_{\disturb  \disturb}  \gain_{\disturb} \right\} \nonumber
	\end{align}
	\label{eq:cost_predict}
\end{subequations}
with the caret symbols signifying predictions upon application of the policies \eqref{eq:feedback_control}.
\section{Computational Procedure and Discussion.}
\label{sec:procedure}

Suppose that the position of points, $\{g.\state_i\}_{i=1}^{N}$, where all trajectories emerge is known. In what follows, we describe the numerical scheme for carrying out the integrations \eqref{eq:cost_predict} and computing the overapproximated level set of $\valuefunc(\state_r+\delta \state, t)$ \ie  \eqref{eq:variational_approx_fwd}.  % into the  value function

\subsection{Computational Scheme}
First, a schedule of controls $\{\control_r(t), \,\disturb_r(t)\}_{t=-T}^{0}$ needed for computing nominal states $\{\state_r(t)\}_{t=-T}^{0}$ is initialized as  a problem-dependent parameter and then used to run $\{\state_r(t)\}_{t=-T}^{0}$; if the nominal control schedules are not available, they can be set from the system's passive dynamics. The predicted cost improvement starting from the final time and going backwards in time is (\cf \eqref{eq:cost_predict}):
\begin{align}
	\mid\hat{\valuefunc}^\ast(\state, \tau) \mid =\int_{0}^{\tau} \min\{\bm{0}, \hamfunc
	- \hamfunc(t; \state_r, \control_r, \disturb_r, \hat{\valuefunc}_{\state}^\ast) \}, \, \tau \gg -T,
	\label{eq:Vnear}
\end{align} 
while the actual cost improvement is given by \eqref{eq:cost_diff}.
%
%\begin{align}
%	\costdiff^\star(\state_r, t) =\valuefunc^\star(\state_r, t) -\valuefunc_r(\state_r, t).
%\end{align}
%
Similar to \cite{JacobsonMayne}, define a cost improvement criterion $\rho>0$ such that,
\begin{align}
	{\costdiff^\star(\state_r, t)}/{\mid\hat{\valuefunc}^\ast(\state_r, t) \mid}>\rho
	\label{eq:stopping_crit}
\end{align} 
 determines the ``closeness" of the cost improvement to the cost prediction. % \ie how far away $\state_r$. 

\begin{algorithm}[tb!]
	\caption{Successive Approximation Scheme.} %
		\label{alg:succ_approx}
	\begin{algorithmic}[1]
		\Procedure{VarHJIInt}{$\eta$ $\rho$}
		\Comment{Given $\eta>0$, $\rho \in (0, 1]$.}
		\Comment{Stop conditions.}
		\State Initialize Buffer $V_{buf} = \emptyset$.
		\ForAll{initial trajectories  $X=\{\bm{\state}_i(t_0)\}_{i=1}^{N} \in \openset$}
		\State Generate schedule $\pi_{r_i}=\{\control_{r_i}(t), \,\disturb_{r_i}(t)\}_{t=0}^{t=-T}$;
		\State Initialize $\mid\hat{\valuefunc}_i(\state_{r_i},t) \mid=\infty$;
		\Comment{\cf \eqref{eq:stopping_crit};}
		\State $\state_{r_i}, \pi_i^\ast$ = \texttt{BackwardPass($\state_i, \pi_{r_i}, \mid\hat{\valuefunc}_i(\cdot)\mid$)};
		\State $\valuefunc_i^\ast(\state_{r_i}+\delta \state_i, t), \, \pi_{r_i}^\ast$= \texttt{ForwardPass($\state_{r_i},\pi_i^\ast)$};% \{\control_i^\ast(t), \,\disturb_i^\ast(t)\}$)}
		\State $\valuefunc_{buf} \leftarrow \valuefunc_{buf} \cup \valuefunc_i^\ast(\state_{r_i}+\delta \state_i, t)$;
		%\Comment{Union with $\valuefunc_{buf}$.}
		\label{line:buffer}
		\EndFor
		\State Compute zero-levelset of $V_{buf}$ 
		\Comment{Using \cite{LevelSetPy}.}
		\label{line:zerolevset}
		\EndProcedure
		\end{algorithmic}
		\hrule
		%\caption{Backward Integration.} %
		\label{alg:bkwd_pass}
	\begin{algorithmic}[1]
		\Function{\texttt{BackwardPass}}{$\state_{i}, \pi_{r_i}, \mid\hat{\valuefunc}_i^\ast \mid$} \label{func:bkwdpass}
		\State Initialize $t_{eff}=-T, \, k=1\cdots K$.
		%\While{$\mid\hat{\valuefunc}_i^\ast(\state, \tau) \mid>\eta$}
		\For{$t=0, -k\Delta t ,\cdots -T$} %\Comment{steps $k\Delta t$} 
		\label{line:bkwdstart}
		\Comment{$\Delta t = T/(K-1)$};
		\State{Unpack $\control_{r_i}(t), \,\disturb_{r_i}(t):=\pi_{r_i}(t)$;}
		\State  $\dot{\state}_{r_i}(t)\leftarrow f(\state_{r_i}, \control_{r_i}(t), \,\disturb_{r_i}(t))$ \& compute $\valuefunc_{r_i}(\state_{r_i}, t)$;
		\State Compute $\mid\hat{\valuefunc}_i^\ast(\state, t)\mid\leftarrow\valuefunc_i^\star(\state_{r_i}, t) -\valuefunc_{r_i}(\state_{r_i}, t)$; %\cf \eqref{eq:cost_diff}
		\If{$\mid\hat{\valuefunc}_i^\ast(\state_{r_i}, t) \mid < \eta$}
		\State  Update $\state_{r_i}$%, set $t_f = t_{ext}$ 
		\Comment{e.g. via Runge-Kutta integration};
		\State Accept $\state_{r_i}$ only if condition \eqref{eq:stopping_crit} is satisfied; 
		\State Set $t_{eff} = t$ and \texttt{Terminate loop}.
		\label{line:smnalldelx}
		\Else{ Line search for a smaller $\delta \state_{r_i}$; restart line \ref{line:bkwdstart}}.
		\EndIf
		\State  $\delta \pi^\ast_i(t):=(\delta \control_{i}^\ast$, $\delta \disturb_{i}^\ast) \leftarrow$ extrema of $\hamfunc_i(\state_{r_i}, \pi^\ast_{r_i}, p, t)$ %in \eqref{eq:gains}.
		\State{Update $\pi_i^\ast(t):=(\control_{i}^\ast(t), \,\disturb_{i}^\ast(t)) \leftarrow \pi_{r_i}(t) + \delta \pi^\ast_i(t)$};
		\EndFor
		\State \Return $\state_{r_i}, \pi_i^\ast :=\{\control_{i}^\ast(t), \,\disturb_{i}^\ast(t)\}_{t=0}^{t=t_{eff}}$.
		\EndFunction
	\end{algorithmic}
	\hrule
	\label{alg:fwd_pass}
	\begin{algorithmic}[1]
		\Function{\texttt{ForwardPass}}{$\state_{r_i}, \pi_i^\ast$}
			\For{$t=-T, (K-1)\Delta t, \cdots, -2 \Delta t, -\Delta t,  0$}
				\State{Unpack controllers $(\control_{i}^\ast(t), \,\disturb_{i}^\ast(t)) := \pi_i^\ast(t)$};
				\State Compute the extremizing $\delta \control\,  \& \, \delta \disturb$ \cf \eqref{eq:rob_rhs}, \eqref{eq:gains};
				%\State Updates $\control_{i}(t), \disturb_{i}(t)\leftarrow  \eqref{eq:feedback_control}$;
				\State Run $\delta \dot{\state}_i(t)\leftarrow f(\cdot, \delta \control_{i}(t), \,\delta\disturb_{i}(t))$;
				\State  $\delta \state_{i}(t) \leftarrow \delta \dot{\state}_{r_i}(t)$ set $t_f = t_{ext}$ %, set $t_f = t_{ext}$
				\Comment{e.g. Euler/RK integration};
				\State Compute $\valuefunc_i^\ast(\state_{r_i}+\delta \state_i, t)$ \cf \eqref{eq:volterra_expand}; 
				\State Update $\pi_{r_i}^\ast(t) := \left(\control_r(t), \, \disturb_r(t)\right)$
				\Comment{Using \eqref{eq:bkwd_controls}};
			\EndFor
			\State \Return $\valuefunc_i^\ast(\state_{r_i}+\delta \state_i, 0), \, \pi_{r_i}^\ast$.
		\EndFunction
\end{algorithmic}
\end{algorithm}

As seen in Algorithm \ref{alg:succ_approx}, the procedure proceeds in two passes for all initial trajectories: 
\begin{inparaenum}[(i)]
	\item in a backward pass, costs \eqref{eq:cost_predict} are estimated with the open loop sequence $\{\control_r(t), \,\disturb_r(t)\}_{0}^{t=-T}$. We generate $\control^\ast$ and $\disturb^\ast$ in \eqref{eq:bkwd_controls} afterwards;
	\item in a forward pass, reversing the order of integration limits, controls $\control^\ast(t), \,\disturb^\ast(t), \,\,t \in \left[-T, 0\right]$ of \eqref{eq:fwd_controls} are then applied and the approximation in \eqref{eq:volterra_expand} is computed for every trajectory that emanates from the state space. 
\end{inparaenum}
%We compute the zero sublevel set of the terminal cost using~\cite{Mitchell2005} overapproximation level set methods, implemented in the author's GPU library~\cite{LevelSetPy}; we compare this zero sublevel set with the  overapproximation coverage of our second order method. The result is presented in \autoref{fig:level_cmp}.
For the line search procedure, an Armijo-Goldstein condition in a typical backtracking line search can be applied to iteratively keep  $\delta \state$ small for a valid approximation of $\valuefunc(\cdot)$ \ie \eqref{eq:variational_approx}. A regularization scheme similar to our previous work~\cite{iDG} can also be applied to keep the stagewise Hessians positive definite. 

The union operator from line \ref{line:buffer} of Algorithm \ref{alg:succ_approx} allows the recovery of the maximal BRS as proposed in \cite{DecompChenHerbert}. Given the ``safe" and ``unsafe" sets that constitute $\valuefunc_{buf}$, the reachable sets can be over-approximated  under a best-response strategy of the two-player game (See~\cite{iDG}). We have released code that computes the zero isocontour (levelset) of the (union of all trajectories') optimal value function stipulated on Line \ref{line:zerolevset} of Algorithm \autoref{alg:succ_approx}. A method for obtaining this is available in~\cite{LevelSetPy}, which is an implementation of~\cite{lorensen1987marching}.
\subsection{Computational Complexity and Convergence Requirement}
The problem introduced in \eqref{eq:variational_approx} can be solved with Newton's method to find the extremizing policies $\delta \pi$ for all $t$ in Algorithm \ref{alg:succ_approx}: with Hessian $\frac{\partial^2 V}{\partial \pi_u^2}$ for the maximizer and $\frac{\partial^2 V}{\partial u^2}, \frac{\partial^2 V}{\partial v^2}$ for the minimizer, the inversion of the Hessian matrix would constitute $O(T^3m^3)$ CPU flops. Whereas the overall DDP-style computational scheme we have presented has a CPU cost per iteration of $O(N) + T \cdot  (2n^3+\frac{7}{2}n^2(n_u+n_v)+2n(n_u+n_v)^2+\frac{1}{3}(n_u+n_v)^3+O(n^2)+O((n_u+n_v)^2))$ CPU flops (see \cite[Appendix II]{Liao1991Convergence} for details), where $n_u$ and $n_v$ are as given in section \ref{sec:notations}. 	The polynomial  time complexity of the presented scheme makes it more attractive compared to Newton's method or even level set methods which are well-known to scale exponentially. We refer readers to Pantoja~\cite{Pantoja} for a thorough differentiation between DDP and Newton's methods. In our opinion, recent first-order primal-dual algorithms such as Chambolle-Pock~\cite{ChambollePock}  may prove more computationally parsimonious for these problems.

\noindent \textbf{A Note on Convergence}: Conditions upon which the algorithm \ref{alg:succ_approx} converges is premised on the standard Hessians' positive-definiteness (PD) \ie $\hamfunc_{\control  \control}$ and $\hamfunc_{\disturb  \disturb}$ of DDP algorithms. In addition, for linear dynamical equations \eqref{eq:sys_dyn}, the cost function being PD convex is a sufficient requirement for PD stagewise Hessians. When ~\ref{eq:sys_dyn} is nonlinear, stagewise PD of $\hamfunc_{\control  \control}$ and $\hamfunc_{\disturb  \disturb}$ is no longer guaranteed~\cite{Shoemaker1991}. In such situations, one may explore \begin{inparaenum}[(i)]
	\item a Levenberg-Marquardt scheme, convert the scheme to steepest descent by turning the stagewise Hessian to an identity or using the \textit{active shift} method of ~\cite{Shoemaker1991}'s  Theorem IV.
\end{inparaenum}

%\addtolength{\textheight}{-12cm}   % This command serves to balance
%\providecommand\BIBentryALTinterwordstretchfactor{2.5}
\bibliographystyle{IEEEtran}
\bibliography{../biblio}	
		
\end{document}